# The differential aging of inertial and non-inertial observers: The Eyewitness observations of a relativistic polygon traveler

J. West

Department of Chemistry and Physics, Indiana State University, Terre Haute, IN 47809

**Abstract:**
The time dilation of non-inertial travelers in circular and polygonal closed paths are well known.  In both cases observers completing a round trip will age less than an observer at rest with respect to the circle / polygon.  This rapid aging is contrary to the slow aging that would be documented by "temporarily comoving" inertial observers who might also move along each edge of the polygon.  A detailed description of what the non-inertial observer might actually see that would explain their own slow aging is presented.  It is also argued that the non-inertial observer will disagree with the comoving inertial observers about the size of the diameter of the polygon.  In particular, the non-inertial traveler will observer a Lorentz-like contraction of distances measured perpendicular to the direction of relative motion.  As a result, the traveler will obtain a value of $\pi$ for the ratio of circumference to diameter, in contradiction to recently published work that predicts a measured ratio less than $\pi$ and a non-Euclidean geometry for the polygon traveler.

# I. INTRODUCTION

Special relativity, with its seemingly paradoxical predictions of slow moving clocks, length contraction, issues of velocity addition, etc., is a challenging topic. While the mathematics involved is not as advanced as that found in many areas of physics, the issues of clock synchronization, multiple frames of reference, along with the introduction of Minkowski diagrams, and four-vectors makes the mental gymnastics of melding physical intuition with the mathematics a nontrivial enterprise. In order to build some physical intuition, the conceptual introduction of light clocks, train cars, and multiple comoving friendly observers with synchronized clocks and meter sticks spread throughout one's reference frame are especially useful. The analysis presented in this paper should be accessible to undergraduate physics majors and provide useful additional conceptual support in understanding the classic "twin" paradox as well as accelerated observers in rotating frames.

Given the wonderful conceptual model provided by light clocks distributed among many observers in inertial frames, it is somewhat surprising the level of difficulty and uncertainty that is introduced when attempting to extend their use and special relativity to cases of accelerated observers.[1-6] The twin "paradox" is the prime example, with a great deal of work expanding on what at first seems to be an extremely simple example. An observer stationed on a rotating disk/ring/cylinder that rotates with a constant angular velocity provides a similar well-defined test case for dealing with observations made by accelerating observers and the Ehrenfest paradox, the rotating frame's signature equivalent of the twin "paradox." Like the twins, Ehrenfest's paradox can be dealt with at a very basic level using appropriately designed light clocks,[2] but there remain a great number of unresolved issues related to measurements made by rotating observers (an excellent collection of contributed papers from much of the recent research on rotating frames can be found in ref. 3). To go from inertial observers to constantly accelerating observers is a large conceptual leap in special relativity.

One very appealing method to deal with these difficulties is to approximate circular motion by closed path motion around an N-sided regular polygon.[4] As is the case with the classic twin "paradox," the round trip is then composed of a series of inertial stages, interrupted by short periods of acceleration. As N becomes "large," one (hopefully) recovers the case of uniform circular motion. It is not clear, however, the degree to which such a non-inertial observer can rely on comoving inertial observers for measurement information.[5,6]

Consider a regular polygon of N sides (N chosen even). An inertial observer, hereafter referred to as "Poly," is fixed at the center of the polygon. She and her friends have measured the size of the polygon and determined that it has a diameter D, defined as the distance from the midpoint of one segment to the midpoint of the segment on the opposite side of the polygon. The circumference of the polygon is L and L = N*S, where S is the length of each segment of the polygon. A second observer, hereafter referred to as "Octo," (who just happens to be Poly's twin brother) travels along the closed path defined by the edges of the polygon. Poly observes that Octo travels at a constant speed v, along the straight edges of the polygon from vertex to vertex, changing direction at each successive vertex. As such, Octo spends the majority of each round trip as an **inertial** observer. Octo changes direction at each vertex via magnetic fields so that Poly agrees that Octo's speed is always v.[4] In addition, there are N sets of inertial observers, hereafter referred to as "the Commuters," traveling with the same speed v relative to the polygon parallel to each side of the polygon, so that Octo travels along with a different set of inertial observers along each edge.

At the start of a lap, Octo is at a vertex and Octo and each Commuter has 1000 muons in their pocket (and who doesn't?). Poly has left a bag of 1000 muons lying at that same vertex. At the end of the trip around the polygon, Octo agrees with **all** observers that he has more muons in his pocket than remain in Poly's bag. Octo claims that Poly has "a fast clock." This experimental fact is confirmed in particle accelerators every day, on a smaller scale in normal sized labs,[7] and in classic physics instructional movies.[8] In contrast, the Commuters claim that there are more muons in Poly's bag than remain in their own pockets (they claim that Poly has "a slow clock"). If Octo and the Commuters disagree about this basic measurement of time, then how can Octo expect to agree with the Commuters regarding measurements of distance?

Under the assumption that Octo does agree with the Commuters, Semon, et. al. conclude that Octo will agree that the circumference L, of the polygon is Lorentz contracted by the usual relativistic factor of $\gamma = (1 - v^2/c^2)^{1/2}$, so that L' = L/$\gamma$. Octo then also agrees that because their relative motion is perpendicular to the diameter of the polygon (circle for large N) it is **not** contracted so that D' = D. It then follows that Octo sees a non-Euclidean geometry as N becomes large, with L'/D' = $\pi/\gamma$. However, given the obvious disagreement on the aging of Poly's muons, it is the suggestion here that Octo should measure D' on his own!

In section II a concrete conceptual example shows that Octo will agree that Poly ages more quickly than himself. In section III a detailed explanation for why Octo can actually watch Poly age more quickly than himself. The disagreement between Octo can comoving observers leads to a discussion in section IV in which Octo uses a round trip light signal (radar distance) to measure the radius of the polygon, finding that it is also contracted by the same relativistic factor as the circumference so that D' = D/$\gamma$. Octo therefore finds that L'/D' = $\pi$ and he recovers his Euclidean geometry. Finally, the results of the paper are summarized in section V.

## II. COMOVING INERTIAL OBSERVERS DISAGREE

Consider the N-Gon Train Service that runs a train switch hub. The hub is in an inertial frame and is composed of eight track segments, each of length S, in the form of a regular octagon (N = 8). There is an office, also in the shape of an octagon, at the center of the hub with a large clock mounted on the outside wall of each side which is easily visible from the tracks. As measured in the hub frame, the distance from the office to the mid-point of any given track segment is R = D/2. The distance to any given vertex, where the switches are located, is a distance X = 1.08 R, and S = 0.828 R (see Fig. 1).

Octo is a rail inspector for the train service, and as such he constantly travels a set of maintenance rails by hand car, all alone. Octo always makes the left turn at each successive switch. Following ref. 4, the method of turning is via magnetic forces so that Octo's relative speed is always v = 0.8c ($\gamma$ = 5/3) as measured in the frame of the hub. While N = 8 may not qualify for N "large," it is large enough to incorporate the fact that Octo changes track segments on a regular basis and yet not so large as to make visualizing Octo's daily routine difficult.

All day long, sets of inertial observers travel in very long trains along all of the straight segments of the polygon track nearly overlapping with the maintenance track. These observers are referred to as the Commuters and their speed is also v = 0.8c, but they never turn at the vertex switches. They have always been, and will always be inertial observers on their own train in constant straight line motion. A Commuter train therefore constitutes a well defined inertial frame with an associated set of dedicated observers with the usual synchronized clocks and uniform meter sticks.

Octo is always sharing stories and information with the Commuters. In the course of these many conversations, Octo notices that he and the Commuters agree that they are moving with a speed v' = 0.8 c relative to the switches. They also agree that the track segment they are currently traveling along is shorter than Poly claims. According to the Commuters, the length of their track segment is only $S' = S/\gamma = 3/5(0.828 R) = 0.4968 R \sim R/2$ (see Fig. 2). Octo and the Commuters agree that they spend a time $t' = S'/v' = (5/8) (R/c)$ traversing each track segment. Note: Commuters on different tracks do not agree with each other (see fig. 2) about the length or orientation of the segment of other Commuter trains. Poly agrees that the trains move relative to the switches at a speed of v = v' = 0.8c. However, Poly can see from the notes in her log book that each traveler takes a time of $t = S/v = 1.035 (R/c) > t'$ to finish their trip along each track segment.

As follows from the usual light clock and train analysis, or the classic muon-decay experiment discussion the Commuters claim Poly has overestimated the transit time because Poly has meter sticks that are too short by the factor $\gamma$ when they are laid parallel to the track segment.[8-10] Poly explains the difference of opinion by the fact that the Commuters (and Octo) have clocks that run to slowly by the factor $\gamma$. Now, the Commuters **also** claim that Poly's clocks run slowly by a factor $\gamma$, and they have been running slowly forever. The Commuters still predict that Poly's log book will show the shorter transit time because Poly's assistants who are responsible for actually recording the arrival times at the switches do not have their clocks properly synchronized with Poly. So the Commuters say that Poly has a slow clock, but the hub data will consistently overestimate the time required to traverse a track segment. For all N-1 track segments other than their own, Octo has a nonzero velocity relative to themselves for all N-1 of those track segments so the Commuters also say that **Octo** ages slowly for most of his round trip.

Octo **must** object to the assertion that Poly has slow running clocks! When Octo and Poly go home for the holidays (remember, they are twins), everyone can plainly see that Poly has **always** aged **more** than Octo, and Octo is always wanting to sell her is "extra" muons. All physical evidence forces Octo to conclude that Poly's clock runs too **quickly**, not too **slowly**. An even more extreme case results if Octo compares himself with another rail repair worker who travels the maintenance rails around polygon in the opposite direction. By symmetry Poly must see the other maintenance worker age the same as Octo, and yet Octo moves relative to the other maintenance worker the entire time (a complete treatment of that case is given in ref. 11, and an extremely interesting variation is given in ref. 12).

How can Octo and the Commuters disagree about Poly's aging? Octo and the Commuters are comoving inertial observers along each segment of track. They can both directly observe, with the Mark I eyeball, the big clock on the office wall. How could they possibly disagree about what they see directly with their eyes while undergoing identical, overlapping inertial motion? The use of instantaneous comoving inertial observers as stand-ins for accelerating observers for measurements beyond a single point in space-time has already been called into question by previous investigators,[6] but Octo is not accelerating during the track segment. The Commuters are not comoving in an "instantaneous" manner. Still one needs to consider how much agreement there can be between Octo and the Commuters for any measurement.

### III. ROUND TRIP OBSERVERS ARE SPECIAL

The conflict between Octo and the Commuters calls for a direct approach. In particular, what does Octo see if he watches (films) Poly for the duration of one full round trip? The Commuters **must** agree with Octo that Poly's clock **looks** like it is running fast. But the Commuters "know" that this is only an optical illusion caused by the relativistic Doppler effect. A similar effect is well known as part of the standard analysis of the classic twin "paradox."[13,14] When the "traveling" twin is in the process of returning to earth, they see the clocks on earth running quickly, but the other observers in the traveling twin's new inertial reference frame know the clocks on earth run too slowly. The clocks only appear to run too quickly, because the clocks are moving closer with each successive tick, reducing the travel time for the image of the clock. If the twin does not stop at earth on the return trip, then looking back at the earth over their shoulder, the twin would see the earth clocks appear to tick very slowly due to the combination of the clocks actually running slowly, and the fact that the clocks mover further away from the twin with each successive tick.

In a similar manner, while the Commuters are headed towards Poly, her clocks appear to run too quickly, despite the fact that her clocks are "really" running slowly. This is "set right" once the Commuters have left the hub behind and Poly's clock appears to run extra slowly, as the light from the clocks then has to catch up with the Commuters. It is important to notice that in order to **look** at Poly, one does **not** look towards the office. In fact, as has been discussed in detail elsewhere, the **view** of the hub from the trains is one of a **rotated** octagon.[15,16] When one "looks at Poly" one is really looking at Poly's location when that light was emitted. For transverse motion relative to a localized source of light, such as the clocks, the phenomenon of Stellar aberration is encountered. For example, when the Commuters reach the midpoint of their segment of track, they know that Poly is **located** directly to their left, but they **see** Poly at her "retarded" location. The directions are related to each other by

$$\cos(\alpha) = \frac{\beta + \cos(\theta)}{1 + \beta \cos(\theta)}, \qquad (1)$$

where $\theta$ is the angle relative to the track to Poly's "known" position, and $\alpha$ is the angle as measured with a protractor in hand that one needs to look to see Poly.[16,17] For example, when the Commuters are at the midpoint of a track segment, they "know" that the direction to Poly's "current" location is $\theta = 90°$ to their left but Poly has been racing towards them with a speed of 0.8c so the image of Poly is seen by "looking" at the angle $\alpha = 36.9°$, well in front of and slightly to the left of the track segment.

A bit of angular book keeping is in order at this point. According to the Commuters, the polygon is not a regular polygon but a Lorentz contracted version (see fig. 2). As they first enter a track segment, the angle to the office is $\theta = 76.1°$, but it is seen at $\alpha = 29.25°$ (see fig. 3a). Even more important for the discussion at hand, when the Commuters and Octo reach the **end** of a given track segment, $\theta = 103.9°$, to the left and slightly behind them, but it appears that Poly is at $\alpha = 46.1°$, still well in **front** of them (see fig. 3b)! It **appears** that Poly is headed toward Octo and the Commuters for their entire time on the track segment!

The relativistic Doppler effect builds in Poly's slow clock, and the apparent relative motion

$$f' = f * \frac{(1-\beta^2)^{1/2}}{1-\beta\cos(\alpha)}, \quad (2)$$

where f is the rate of Poly's clock as seen by Poly (1 tick/second) and f' is the rate of Poly's clock as filmed by the Commuters and Octo.[18] It is important to note that it is the angle $\alpha$ that is used in Eq. (2), and **not** $\theta$.[19,20]

Based on Eq. (2), the Commuters and Octo agree that it **appears** that Poly's clock is running too quickly during their entire time on the track segment. Using the values of $\alpha$ found via Eq. (1), one can show that at the: start of the segment f' = 1.99f; midpoint f' = 1.66f; end of the track segment f' = 1.35 f. Now, the crux of the matter is that Octo changes to a new track segment at each switch, and for the entirety of every track segment Octo sees Poly's clock run too quickly. Octo is not surprised at all by Poly's advanced age when they get together for vacation. It should be mentioned that it is certainly possible for $\alpha$ to become larger than $90°$ and Octo would then see Poly's clock run slowly for a short time. The values v = 0.8c and N = 8 were purposely chosen large enough to provide an example where Octo sees Poly's clock run too quickly for the entire segment.

### IV. DIAMETER MEASUREMENT: OCTO'S FLAT GEOMETRY

Octo's confidence in the Commuter's ability to make accurate measurements is squashed, and Octo resolves to make his own measurements whenever possible. In particular, Octo would like to measure R. Consider letting N become "large," so that the polygon approximates a circle. Octo will use "radar time" to determine the value of R. He sets off a very bright flashbulb, and then simply counts off the seconds until he **sees** Poly's office brighten.[21,22] Aiming will not be much of an issue, because for N large, $\alpha$ is nearly constant. For uniform circular motion, Poly does not appear to move at all![2] Poly and her associates in the hub easily determine that the round trip time for the flash is t = 2R/c = D/c, and Octo has completed approximately N/4 turns during that time. Octo and Poly both agree that Octo's watch runs slowly compared to Poly's clock by the factor $\gamma$, so Octo will measure a round trip time for the flash of t' = $2R/\gamma c$. Octo then calculates using his own flash and watch a diameter of D' = ct' = $D/\gamma$. It has already been determined that Octo believes that the circumference of the track is L' = $\pi D/\gamma$, so that L'/D' = $\pi$, contrary to the conclusions reached in ref. 4. This does presume that Octo still considers the speed of light to be equal to c along the radial direction, a presumption that is not out of necessity true.[2]

### V. CONCLUSIONS

The use of temporary comoving inertial observers to derive the expected results of the measurement of distances by an accelerated observer would seem to be a natural and logical approach. This would seem to be even more certain in the case of an observer who accelerates only briefly at the vertices of a closed path around a regular polygon. It has been demonstrated, however, that there are difficulties and subtleties associated with pairing measurements made by truly inertial observers, and those made by an observer traversing such a polygon. This is despite the fact that the polygon path allows for inertial motion for almost the entire round trip. The well known and experimentally confirmed result that a traveler making a trip completely around the polygon will have aged less than inertial observers at rest with respect to the polygon is directly explained by the relativistic Doppler effect and the direct observations of the traveling observer. The use of the polygon path is extremely useful in the analysis. For example, it is not clear to the author how such an analysis would be possible for circular motion, where residents at the center of the circle do not have any apparent motion.

In addition to measurements of the clock rate of inertial observers at rest with respect to the polygon, the use of radar time to make distance measurements suggest that the accelerated traveler sees a Lorentz –style contraction of distances perpendicular to the direction of motion. An observer in uniform circular motion would therefore not believe the geometry is non-Euclidean. As noted in ref. 6, the results presented here suggest that extreme caution must be used when utilizing "instantaneously local inertial frames" in the analysis of accelerated systems.

**ACKNOWLEGEMENTS**
The author would like to thank Michelle Baltz-Knorr (ISU) for help in editing the manuscript.

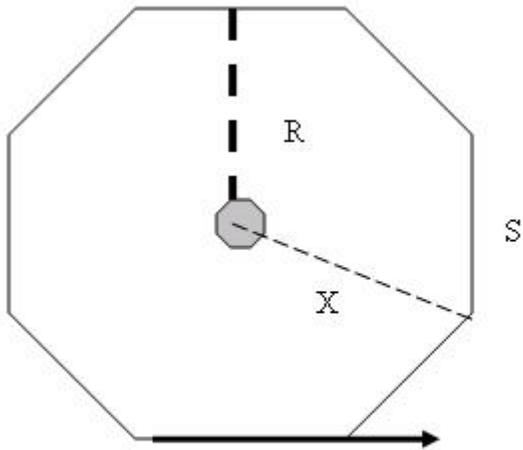

Figure 1. The track layout according to an inertial observer in the frame of the train yard. R is the distance from the station to the center of each track segment (heavy dashed line), X = 1.08R is the distance from the station to each vertex (thin dashed line), and S = 0.828R is the length of each track segment. The heavy arrow shows the direction of travel for a train along the bottom track segment. Trains always travel at a speed of 0.8c in a counter-clockwise direction.

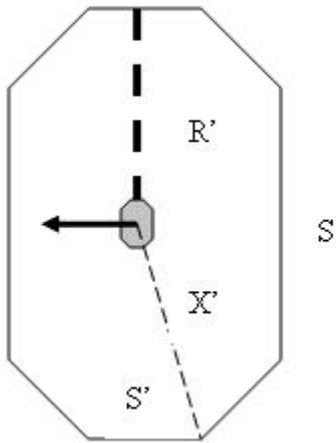

Figure 2. The train yard in the inertial frame of the Commuters. The entire train yard moves to the left at a constant speed of 0.8c. Notice that their segment is Lorentz contracted (S' = S/$\gamma$ ~ R/2), but the radius is not (R' = R).

3a.

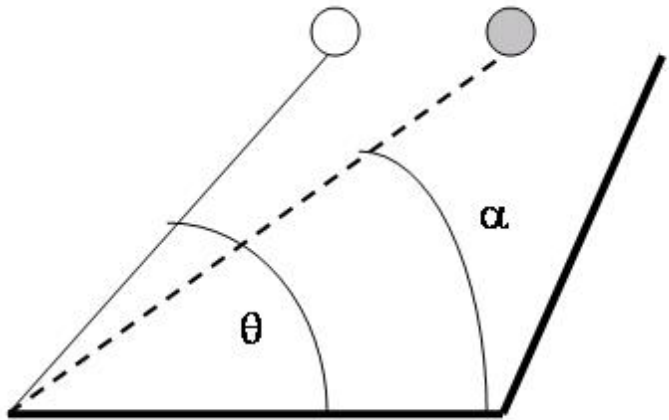

3b.

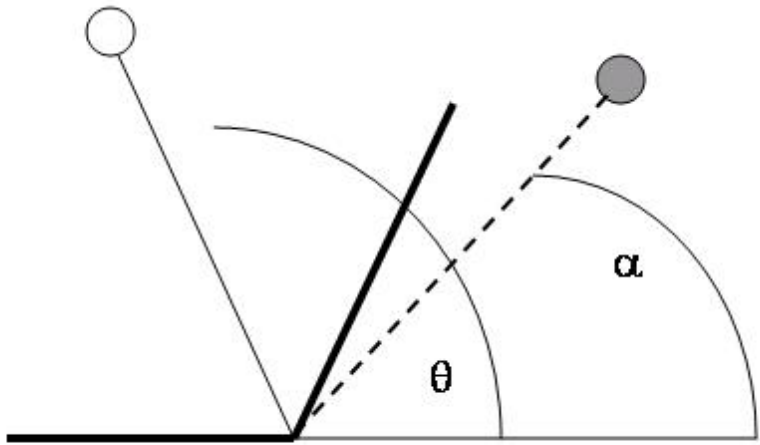

Figure 3. Due to the effects of Stellar aberration, Poly's position and Poly's image location do not coincide. The position of the station (Poly) as measured by the Commuters (angle $\theta$, solid line and empty circle) and where the Commuters actually look in order to see Poly (angle $\alpha$, dashed line and solid circle). 3a) Shows Octo's view at the start of a track segment ($\theta = 76.1°$ and $\alpha = 29.25°$), and 3b) shows his view at the end of a track segment ($\theta = 103.9°$ and $\alpha = 46.1°$). Octo will see the scene reset to view 3a) upon switching to the next segment of track.